# Lagarto I – Una plataforma hardware/software de arquitectura de computadoras para la academia e investigación

Cristóbal Ramírez, César Hernández, Carlos Rojas Morales,
Gustavo Mondragón García, Luis A. Villa, Marco A. Ramírez

Instituto Politécnico Nacional, Centro de investigación en Computación,
México

ralc_@hotmail.com,{hdzces,carlosrojo3000}@gmail.com,
gustavomog@hotmail.com, {lvilla, mars}@cic.ipn.mx

**Resumen.** En la actualidad, la enseñanza del diseño de arquitectura de procesadores sigue siendo una tarea compleja, debido a la cantidad de información que se maneja y a la falta de herramientas enfocadas a condensar esta información, haciéndola más digerible para el estudiante. En este sentido, el grupo de arquitectura de computadoras del centro de Investigación en Computación del Instituto Politécnico Nacional trabaja en un proyecto llamado "Lagarto" con el objetivo de generar una plataforma de cómputo abierta para la academia e investigación, que facilite la comprensión de conceptos fundamentales de Arquitectura de Computadoras y Sistemas Operativos. Este trabajo presenta la arquitectura de un núcleo llamada Lagarto I, el cual es un procesador escalar segmentado de 32-bits que ejecuta el set de instrucciones MIPS32 R6 para operaciones de tipo entero y punto flotante. La arquitectura está descrita utilizando Verilog HDL y es sintetizable en un FPGA. Así mismo, se abordan las diferentes maneras en que es posible probar la arquitectura con códigos escritos en lenguaje C o ensamblador.

**Palabras clave:** arquitectura de computadoras, diseño de procesadores, compiladores, lenguaje ensamblador, FPGA.

## Lagarto I - A Computer Architecture Hardware/Software Platform for Academia and Research

**Abstract:** Nowadays, the teaching of processor architecture design continues to be a complex task, due to the amount of information that is handled and the lack of tools focused on condensing this information, making it easier to understand for the student. In this sense, the computer architecture research team at the Centro de Investigación en Computación of the Instituto Politécnico Nacional is working on a project called "Lagarto" with the aim of generating an open computing platform for academia and research to ease the understanding of fundamental concepts of Computer Architecture and Operating Systems. This paper presents the architecture of the Lagarto I core, a 32-bit pipelined scalar processor that executes the MIPS32 R6 instruction set for integer and floating-point operations. The architecture is described using Verilog HDL and is synthesizable in an FPGA. Likewise, the different ways in which it is possible to test the architecture, with codes written in C or assembly language, are addressed.

## 1. Introducción

Hoy en día, el curso de Arquitectura de Computadoras sigue siendo uno de los cursos fundamentales en la formación de estudiantes de Ciencias de la Computación o Ingeniería de Cómputo. Comprender el comportamiento de un procesador durante la ejecución de un programa no es una tarea fácil, desafortunadamente, a lo largo del curso, los estudiantes se enfrentan con la necesidad de invertir una gran cantidad de tiempo intentando comprender conceptos abstractos, que pueden ir desde la codificación de un programa en alto nivel, pasando por las etapas de compilación hasta llegar a código máquina, terminando con la ejecución de instrucciones en el *pipeline* de un procesador, entorpeciendo el proceso de asimilación de los contenidos del curso, y por consiguiente, no llegar a cumplir con los objetivos planteados. Una solución a esta problemática, es la de facilitar al estudiante una herramienta que permita visualizar el flujo de información a lo largo de las etapas del *pipeline* del procesador: señales de control decodificadas, error en la predicción de saltos, errores de cache, entre otros.

En este sentido, se han propuesto diferentes metodologías para la enseñanza de Arquitectura de Computadoras, partiendo del hecho que la enseñanza se vuelve ineficiente si los métodos son enfocados únicamente en aspectos teóricos. Algunas propuestas plantean el uso de simuladores de arquitecturas de computadoras con el fin de comprender los conocimientos teóricos con una experiencia práctica. Sin embargo, por experiencia sabemos que esta práctica no cumple con el objetivo fundamental del curso: el diseño de procesadores; ya que, desde la perspectiva del estudiante, la brecha que existe entre lo visto en clase y un procesador real puede ser muy grande.

Para afrontar este problema, universidades como UC Berkeley y el MIT han desarrollado una plataforma abierta llamada RISC-V [3], que originalmente surge para fortalecer los conceptos fundamentales de arquitectura de computadoras tanto en la enseñanza como en la investigación, y que ha llevado al desarrollo de diferentes arquitecturas [4]. Actualmente, departamentos de arquitectura de computadores de diferentes universidades como lo son la Universidad de Manchester y la Universidad Politécnica de Cataluña, han hecho uso de esta plataforma para la impartición de cursos y realizar investigación sobre arquitecturas multi-núcleos, NoC, etc. LEON es otra propuesta de un procesador de 32-bit que ejecuta el ISA SPARK V8, desarrollada en VHDL. El modelo es altamente reconfigurable y enfocado para diseños de *System-on-*

*Chip* (SoC). El código fuente de este procesador se encuentra disponible y con uso ilimitado para investigación y educación [5, 6].

En las referencias antes mencionadas se observa que, el proceso pedagógico en la enseñanza de Arquitectura de Computadoras parte desde un supuesto en donde el alcance del curso no es la implementación física del diseño del procesador. Se omite todo el conocimiento que se genera al llevar cualquier modelo teórico-simulado al mundo real. En un mundo en donde el desarrollo tecnológico y la innovación marcan la diferencia en el desarrollo de la sociedad, llegar hasta la última milla, en donde la idea se convierte en un producto tangible, proporciona el conocimiento que conduce hacia una sociedad emprendedora.

En este trabajo se describe el núcleo de un SoC de código abierto llamado Lagarto I [7], desarrollado para facilitar la comprensión del diseño arquitectural y el comportamiento de un procesador RISC de 32-bit segmentado desde diferentes perspectivas: Diseño, Implementación, Simulación y Validación.

El resto del documento está organizado de la siguiente manera. En la Sección 2 se describe la microarquitectura del procesador Lagarto I. La Sección 3 discute las diferentes maneras en las que se puede usar y probar la arquitectura. La Sección 4 muestra los resultados obtenidos al ejecutar diferentes programas en el procesador. Finalmente, las conclusiones y trabajo futuro son presentados en la Sección 5.

## 2. Microarquitectura del procesador Lagarto I

Desde una perspectiva general, Lagarto I puede ser visto como un procesador de 5 etapas (Fig. 1):

— Búsqueda y extracción de instrucciones,
— Decodificación,
— Lectura de registros/Emisión,
— Ejecución,
— Escritura de Retorno.

**Tabla 1.** Unidades de ejecución / etapas.

| Unidad de ejecución | Etapas |
|---|---|
| Enteros | 1 |
| Load/Store | 2 |
| Branch | 1 |
| FP – Operaciones Simples | 1 |
| FP – Operaciones Complejas | 4 |
| FP – Operaciones Complejas | 12 |

Sin embargo, a partir de la etapa de Lectura de registros la instrucción es emitida a su unidad de ejecución correspondiente, siempre y cuando ésta cumpla con los requerimientos para ser emitida. Se tienen diferentes unidades de ejecución con diferente latencia dentro de la arquitectura, tal y como se muestra en la Tabla 1.

El diseño contempla unidades de ejecución de Punto Flotante de alto rendimiento [8] que cumplen con el estándar IEEE 754 [9].

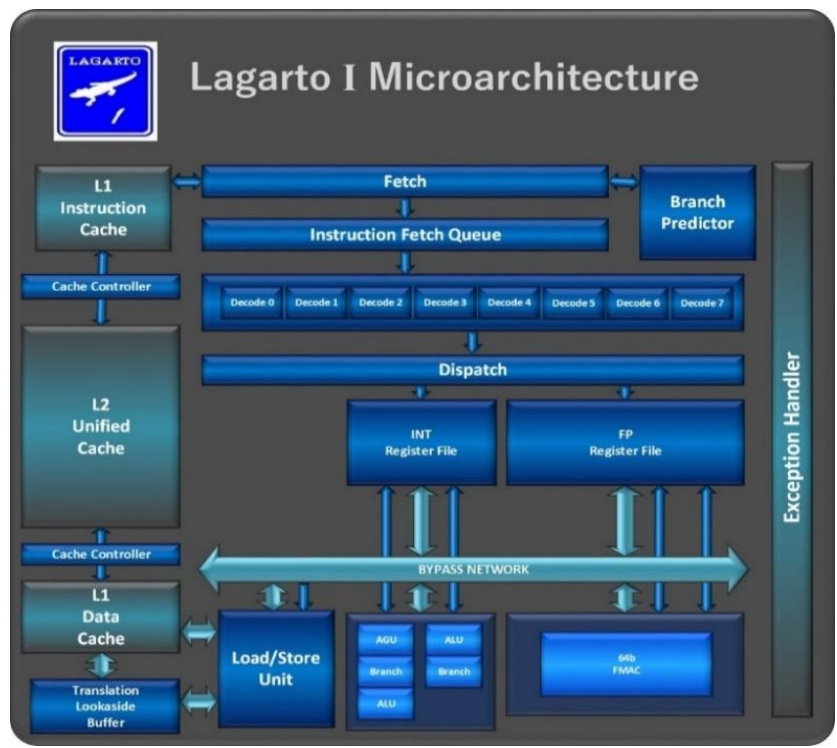

**Fig. 1.** Microarquitectura del procesador Lagarto I.

Lagarto I hace uso de un predictor dinámico y una red de *bypass* completa para adelantar datos entre diferentes unidades funcionales: unidad de enteros, unidad de punto flotante, unidad de load/store, entre otras, mejorando el rendimiento del procesador.

El diseño modular del procesador Lagarto I, permite al estudiante intercambiar diferentes bloques, pudiendo haber sido desarrollados por ellos mismos, como lo pueden ser diferentes predictores de saltos para evaluar rendimiento de cada uno de ellos, aumentar o disminuir los tamaños de las memorias caché.

## 3. Formas de usar y probar Lagarto I

Existen diferentes maneras de usar y probar Lagarto I. La Fig. 2 muestra en la parte izquierda las formas en las que se puede escribir código, ya sea usando un compilador de "C", o bien directamente en código ensamblador. Ambos códigos son traducidos a código máquina, siendo éste el que es interpretado por el procesador (unos y ceros), y dichos códigos son cargados a las memorias caché del procesador para así poder ser ejecutados; del lado derecho se muestran dos interfaces en las cuales se pueden ver los resultados que arroja el procesador.

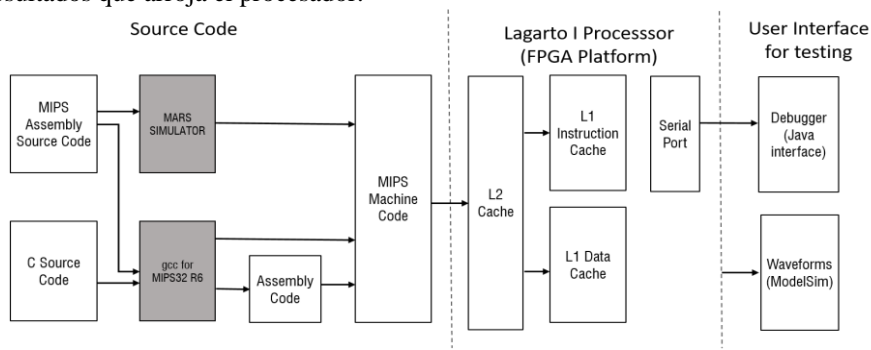

**Fig. 2.** Entorno de uso de Lagarto I.

### 3.1 Programando en lenguaje Ensamblador

Para escribir código en MIPS ensamblador para Lagarto I, es utilizado el simulador MARS, desarrollado en la Universidad Estatal de Missouri; este es un entorno interactivo de desarrollo (IDE) para programación en lenguaje ensamblador MIPS y enfocado para la educación [10].

### 3.2 Programando en lenguaje C

Una segunda opción de generar código máquina para Lagarto I, es escribir código en lenguaje C y compilarlo para arquitecturas MIPS. Lagarto I utiliza el ISA MIPS32 R6, por lo que el compilador debe soportar dicho ISA que corresponde al más actualizado a la fecha. Codescape MIPS SDK [11] provee un compilador completo y un *toolchain* para desarrollo MIPS.

### 3.3 Core Simulation

Es posible utilizar el *soft-core* Lagarto I sin necesidad de ser sintetizado en un FPGA al simularlo en ModelSim, un entorno de simulación multi-lenguaje HDL. Dentro de este entorno es posible correr un banco de pruebas y utilizar una plantilla personalizada (archivo *.do*) para visualizar cada una de las etapas del procesador.

### 3.4 Depurador Lagarto I

Depurador Lagarto I es una interfaz gráfica para el usuario, desarrollada en el Centro de Investigación en Computación del Instituto Politécnico Nacional, la cual permite visualizar los datos computados dentro del procesador Lagarto I cuando se encuentra sintetizado en un FPGA. Esta aplicación cuenta con diferentes características como lo son: control de ejecución paso a paso, control para disminuir o aumentar la frecuencia de operación del procesador, permite modificar el contenido de las memorias cache mediante la carga de nuevos programas o datos, entre otras.

## 4. Pruebas y resultados

Para verificar el correcto funcionamiento del procesador, se codifica un programa en lenguaje ensamblador que calcula el valor de PI mediante el algoritmo de Chudnovsky, el cual es un método rápido para calcular los dígitos de PI. En 2011 se calcularon los primeros 10 billones de dígitos haciendo uso de este algoritmo. Dicho algoritmo es elegido debido a la variedad de las operaciones aritméticas de las que hace uso, así como de operaciones de conversión de entero a punto flotante, movimientos entre bancos de registros, gestionando una gran cantidad de dependencias de datos entre instrucciones, lo que permite comprobar el correcto funcionamiento de distintos módulos de la arquitectura. El programa principal hace llamadas a otras funciones como potencia, raíz cuadrada y factorial, pudiéndose validar el correcto funcionamiento del predictor de saltos y el control implementado en el pipeline del procesador.

Dicho programa fue compilado y ejecutado en el software MARS y el resultado final es mostrado en la Fig. 3, en el registro $f8.

**Fig. 3.** Banco de registros de punto flotante en el software MARS.

En la Fig. 4 se aprecia la conversión del valor de PI tomando en cuenta los primeros 25 decimales al formato de Punto Flotante precisión simple (*float* en lenguaje C), que corresponde al obtenido con el simulador MARS bit a bit (**40490FDA).**

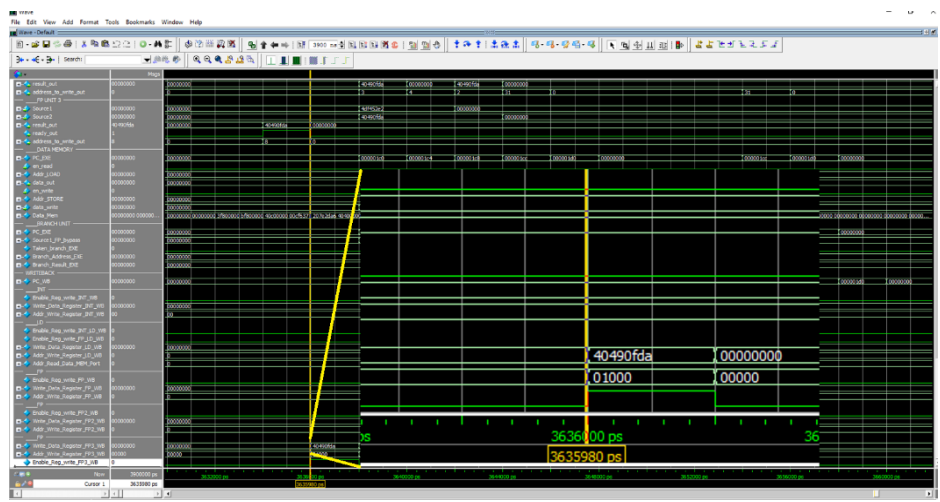

**Fig. 4.** Conversión del valor de PI a formato de Punto Flotante.

En la simulación de la ejecución de este programa sobre Lagarto I se obtienen los resultados mostrados en la Fig. 5, validando el correcto funcionamiento del diseño.

**Fig. 5.** Resultado de la ejecución del programa que realiza el cálculo de PI utilizando el Simulador de Altera ModelSim.

Adicionalmente, el proyecto es sintetizado usando una tarjeta de desarrollo DE2-115 de Altera, e inicializando las memorias de datos e instrucciones con el programa a ejecutar. Los resultados se muestran en la Fig. 6.

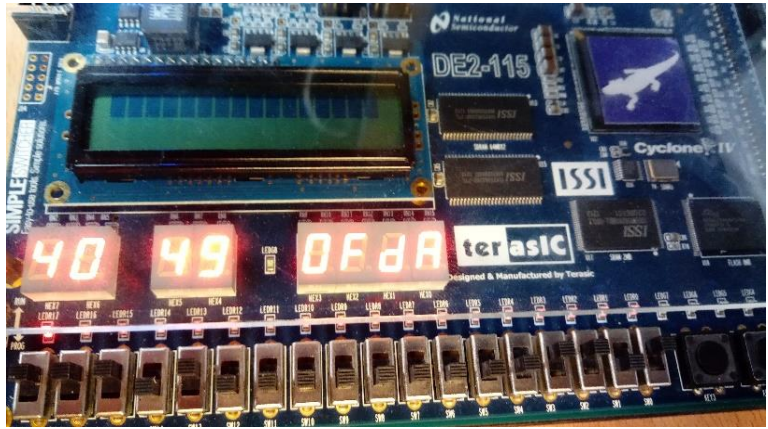

**Fig. 6.** Resultado de la ejecución del programa que realiza el cálculo de PI sobre el procesador Lagarto I sintetizado en el FPGA.

Posteriormente, se visualiza el cómputo del programa paso a paso para monitorear los datos almacenados en el banco de registros y la memoria de datos haciendo uso del Depurador diseñado para la arquitectura Lagarto. El resultado calculado en el FPGA se puede visualizar en la interfaz gráfica, tal y como se muestra en la Fig. 7.

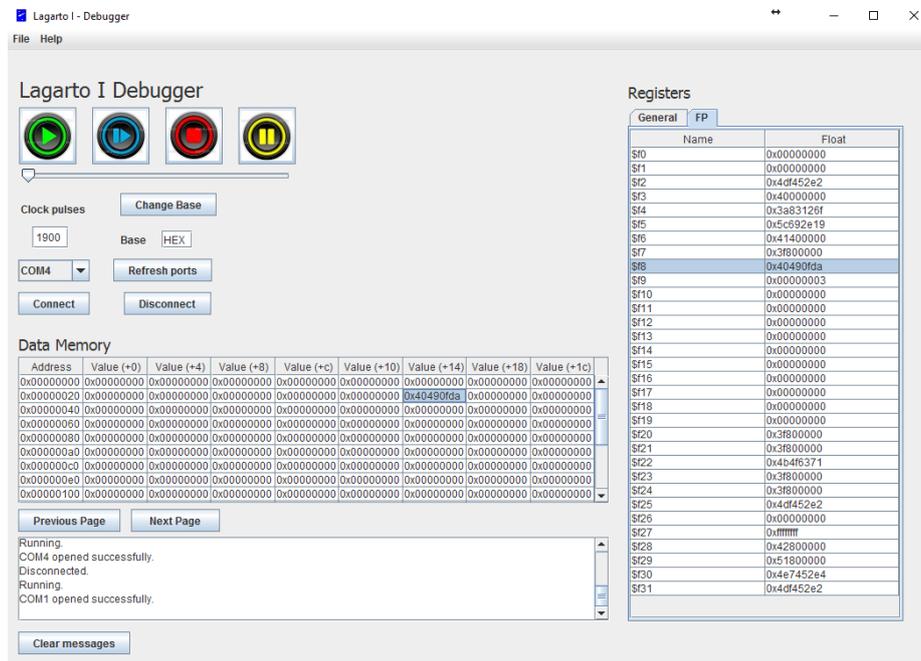

**Fig. 7.** Resultado de la ejecución del programa que realiza el cálculo de PI sobre el procesador Lagarto I sintetizado en el FPGA conectado a la interfaz gráfica.

Por otro lado, el algoritmo es codificado en lenguaje C, y compilado para una arquitectura MIPS, como se muestra en la Fig. 8.

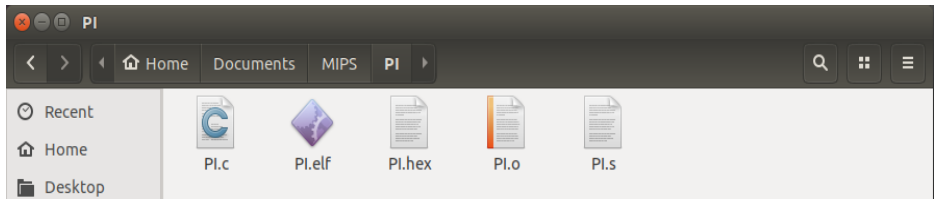

**Fig. 8.** Proceso de Compilación para una arquitectura MIPS.

Se puede apreciar que un procesador de 32 bits logra obtener los primeros 6 dígitos exactos del valor de pi, mientras el resto es una aproximación al valor verdadero.

En la Fig. 9, se muestran los archivos creados en el proceso de compilación en donde primero se creó el archivo objeto (.*o*), posteriormente el archivo (.*elf*) que es un ejecutable para arquitecturas MIPS tal y como se muestra en la Fig. 10. El archivo (.*s*) contiene el programa en ensamblador correspondiente al código escrito en C.

**Fig. 9.** Archivos creados en la compilación para una arquitectura MIPS.

**Fig. 10.** Archivo ejecutable para una arquitectura MIPS.

Finalmente, haciendo uso del comando *hexdump* sobre el ejecutable obtenido al realizar la compilación para MIPS, se obtienen los archivos para inicializar a las memorias caché del procesador y ejecutar el programa en Lagarto I.

## 5.     Conclusiones y trabajo futuro

En este trabajo es presentada la microarquitectura del procesador Lagarto I, una plataforma abierta enfocada en facilitar la comprensión y el uso de procesadores para investigación y la academia, aprovechando los beneficios de los FPGA y los lenguajes HDL. Lagarto I es parte de un *System-on-Chip* (SoC) que se encuentra bajo desarrollo en el centro de Investigación en Computación, y el cual ha abierto las puertas al desarrollo de diferentes proyectos como redes de interconexión, extensiones multimedia, controladores de periféricos, entre otros. Finalmente, Lagarto I como herramienta educativa, está siendo introducida impartiendo talleres en diferentes congresos y universidades en México, despertando el interés por parte de profesores y estudiantes.

## Referencias